\documentclass[12pt,psfig]{article}
\usepackage{graphicx}
\begin{document}

\begin{center} 
{\bf  Predicting the $\Lambda$ binding energy in nuclear matter        
 }              
\end{center}
\vspace{0.1cm} 
\begin{center} 
 F. Sammarruca \\ 
\vspace{0.2cm} 
 Physics Department, University of Idaho, Moscow, ID 83844, U.S.A   
\end{center} 
\begin{abstract}
The purpose of this note is to report  predictions of 
the binding energy of the $\Lambda$ hyperon in nuclear matter
using the latest version of the 
 J{\"u}lich nucleon-hyperon meson-exchange potential.                   
Results from a     
conventional Brueckner calculation are compared with                 
previously reported values. 
A calculation including Dirac effects on the $\Lambda$ single-particle
potential is also presented. Issues encountered in Dirac calculations with
nucleon-hyperon potentials are discussed. 
\\ \\ 
PACS number(s): 21.65.+f,21.80.+a 
\end{abstract}

\section{Introduction} 
                                                                     
There are important motivations for including strange baryons in nuclear matter.
The presence of hyperons in stellar matter tends to 
soften the equation of state (EoS), with the consequence that the predicted             
neutron star maximum masses become considerably smaller. With recent 
constraints allowing maximum masses larger than previously accepted limits 
\cite{Fuchs}, 
accurate microscopic calculations which include strangeness (in addition to other  
important effects, such as those originating from relativity), 
become especially important and timely. 
Furthermore, as far as terrestrial nuclear physics is concerned, studies 
of hyperon energies in nuclear matter naturally complement our knowledge
of hypernuclei.                  

Microscopic calculations of nuclear matter properties have been reported earlier  
within the non-relativistic Brueckner-Hartree-Fock framework (see, for instance, 
Refs.~\cite{catania1,catania2}), using             
the Nijmegen \cite{Nij89} and/or 
the J{\"u}lich \cite{NY89,NY94} nucleon-hyperon (NY) meson-exchange potentials.                                     

In this work we use the most recent meson-exchange NY potential                 
from the  J{\"u}lich group \cite{NY05}.                                                                                    
Given that there are significant differences between this and the previous 
(energy independent) version 
of the J{\"u}lich NY potential \cite{NY94}, it will be interesting to see how those differences reflect 
onto $G$-matrix calculations. Before moving on to our final objective, which is a 
microscopic determination of the EoS for hyperonic matter, we will 
first confront a much simpler scenario, namely the one of nuclear matter at some 
Fermi momentum $k_F^N$ in the presence of a ``$\Lambda$ impurity".                       
This calculation, the outcome of which is the binding energy of 
a $\Lambda$ hyperon in nuclear matter, 
will allow us to explore the behavior of the new NY potential in nuclear 
matter before addressing more involved situations. 
Some empirical information is available for the $\Lambda$ binding energy from 
analyses of $(\pi,K)$ and $(K,\pi)$ reactions and studies of energies of hypernuclei   
\cite{Dover+}. 

We will also report on                                                       
our prediction for the $\Lambda$ binding energy from 
a Dirac-Brueckner-Hartree-Fock (DBHF) calculation.                 
To the best of our knowledge,             
such calculation has not been performed before. 

\section{Calculations using the non-relativistic Brueckner G-matrix} 
For matter with non-vanishing hyperonic density, 
the nucleon, $\Lambda$, and $\Sigma$ single-particle potentials are the solution of a    
coupled self-consistency problem, which reads, schematically 
\begin{eqnarray}
U_N = \int_{k<k_F^N} G_{NN} + \int_{k<k_F^{\Lambda}} G_{N \Lambda}                                       
 \int_{k<k_F^{\Sigma}} G_{N \Sigma}                                       
 \\ \nonumber                                              
U_{\Lambda} = \int_{k<k_F^N} G_{\Lambda N} + \int_{k<k_F^{\Lambda}} G_{\Lambda \Lambda}                             
+ \int_{k<k_F^{\Sigma}} G_{\Lambda \Sigma}                             
 \\ \nonumber                                              
U_{\Sigma} = \int_{k<k_F^N} G_{\Sigma N} + \int_{k<k_F^{\Lambda}} G_{\Sigma \Lambda}                             
+\int_{k<k_F^{\Sigma}} G_{\Sigma \Sigma}                             
\label{fs:eq9} 
\end{eqnarray}
In the equations above, $G_{NN}$, $G_{NY}$, and $G_{YY'}$,                                     
($Y,Y'=\Lambda,\Sigma$), are the nucleon-nucleon,      
nucleon-hyperon, and hyperon-hyperon $G$-matrices at some nucleon and hyperon densities 
defined by the Fermi momenta $k_F^N$ and $k_F^Y$. 

Following the earlier calculation from Ref.~\cite{NY94}, which we want to compare with, 
we make the following 
assumptions and approximations:
\begin{itemize}
\item
We consider the case of symmetric nuclear matter at some 
Fermi momentum $k_F^N$ in the presence of a ``$\Lambda$ impurity", i.e.                  
$k_F^{\Lambda}\approx 0$.                                                
\item 
We take the single-nucleon potential from a separate         
calculation of symmetric matter \cite{AS03}. Notice that the $\Lambda$ potential is
quite insensitive to the choice of $U_N$, as reported in Ref.~\cite{NY94} and as we have 
observed as well. 
\item 
Both $\Lambda$ and $\Sigma$                        
are included in the coupled-channel calculation of the NY $G$-matrix, but              
free-space energies are used for the latter.                         
\end {itemize}
The parameters of the $\Lambda$ potential, on the other hand, are calculated self-consistently
with the $G_{NY}$ interaction, which is the solution of the Bethe-Goldstone 
equation with one-boson exchange NY potentials. In the Brueckner calculation,
density-dependent effects come in through 
angle-averaged Pauli blocking and dispersion.                                      

For some NY initial state, the starting energy that enters in the scattering
equation is 
\begin{equation}
E_0= e_{\Lambda}(p_{\Lambda}) + e_N(p_N).                                 
\end{equation} 
For nucleons and $\Lambda$'s, we write the single-particle energy, in the 
non-relativistic case, as 
\begin{equation}
 e_i(p) = \frac{p^2_i}{2m_i} + U_i(p_i) + m_i 
\end{equation} 
($i=N,\Lambda$). 
(Angle-averaged) Pauli blocking is applied to all intermediate two-baryon states.
The integration in the Bethe-Goldstone equation is handled using standard methods
to eliminate the angular dependence \cite{NY94}. 
We adopt the continuous choice for the single-particle potential. Once the latter is 
obtained, the value of $-U_{\Lambda}(p_{\Lambda})$ at 
$p_{\Lambda}$=0 provides the $\Lambda$ binding energy in nuclear matter, $B_{\Lambda}$. 

In this work, we will apply the latest meson-exchange model by the 
J{\"u}lich group \cite{NY05}, which will be denoted by NY05. 
As shown and discussed extensively in Ref.~\cite{NY05},          
there are several remarkable differences between this model and the previous 
$NY$ J{\"u}lich potential \cite{NY94} (NY94).                  
The main new feature of NY05 is            
a microscopic model of correlated $\pi \pi$ and $K \bar{K}$ exchange to constrain both the 
$\sigma$ and $\rho$ contributions \cite{NY05}. The usual one-boson-exchange contributions
from the lowest pseudoscalar and vector meson multiplets are also present, with the coupling
constants determined by SU(6). This makes the long and intermediate range parts of the potential
well determined.                                   
New short-range features are included through the $a_0(980)$ meson and a strange scalar meson 
with a mass of approximately 1000 MeV. Both of these are parametrized phenomenologically in 
terms of one-boson exchanges in the respective spin-isospin channels. 
The model describes well the available data on integrated as well as differential cross sections
\cite{NY05}. Also, the hypertriton binding energy is well reproduced \cite{triton}. 

There are, though, some major quantitative differences between NY05 and NY94 in specific
partial waves. 
Most noticeably, 
the new model predicts a considerably larger scattering length in the singlet channel. 
These differences turn out to have a large impact on 
in-medium predictions.   
With the new model, we obtain                        
considerably more attraction than Reuber {\it et al.}~\cite{NY94}, approximately            
50 MeV at $k_F^N$=1.35 fm$^{-1}$ for $B_{\Lambda}$, rather than 30 MeV. A value of 49.7 MeV 
has been reported by H. Polinder with NY05 \cite{Polinder}.  
Notice that a value of 30 MeV is generally accepted as the ``empirical" one \cite{Dover+}, 
which opens some interesting questions: 
\begin{itemize}
\item 
The value predicted by NY05 for the  hypertriton binding energy is 2.27 MeV,  
in satisfactory agreement with 
the experimental value of 2.354(50) MeV.                                    
How reliable is the ``empirical" value for the $\Lambda$ binding energy in nuclear matter?                                           
\item 
 How will the additional attraction impact the EoS and 
neutron star predictions? How will those predictions compare to the most                            
recent constraints?                
\end{itemize}

To better highlight the potential model dependence of the predicted   
$\Lambda$ binding energy,                                                       
 we show in Table 1 how selected partial waves contribute to it. 
For each partial wave, column NY94 shows the results given in 
Ref.~\cite{Polinder}, whereas column NY05 are the predictions from this work. 
Clearly the largest contribution to the model
dependence originates from the $S$-waves, although the contribution from the $P$-waves   
is also dramatically different between the two sets of predictions (but much smaller 
than the one from the $S$-waves). 

\begin{table}
\caption{Contributions to the $\Lambda$ binding energy from selected
partial waves as obtained in non-relativistic Brueckner-Hartee-Fock calculations. 
NY05 and NY94 indicate the J{\"u}lich NY potentials from 
Ref.~\cite{NY05} and Ref.~\cite{NY94}, respectively.
} 
\begin{center}
\begin{tabular}{|c|c|c|}
%\hline
Partial wave & NY94, BHF       & NY05, BHF               \\                                         
                  \cline{1-3}
$^1S_0$ &                3.6           & 8.73                       \\ 
                  \cline{1-3}
$^3S_1$+$^3D_1$ &        27.2         & 37.69                      \\ 
                  \cline{1-3}
$^3P_0$   &             -0.6          & 0.69                       \\ 
                  \cline{1-3}
$^1P_1$+$^3P_1$ &        -2.0         & 0.13                       \\ 
                  \cline{1-3}
$^3P_2$+$^3F_2$ &       0.8           & 3.27                       \\ 
\cline{1-3}
Total (all states)  &                29.8          & 51.27                      \\ 

%\hline
\end{tabular}
\end{center}
\end{table}
The NY05 entries agree well with those from Ref.~\cite{Polinder} for the same potential; the relatively
minor differences are possibly due to different choices for the single-particle
spectrum, the handling of the angular dependence in the scattering equation, and, to a very small 
extent, the choice of the nucleon potential.                                      

\section{Dirac effects on the $\Lambda$ binding energy}                                
The relation between the non-relativistic Brueckner approach and the relativistic
framework (known as Dirac-Brueckner-Hartree-Fock, DBHF)                                 
has been discussed for a long time. Already in Ref.~\cite{Brown} it was
shown how relativistic effects tie in with virtual excitations of pair terms. 
Lately, these concepts have been revisited in more detail \cite{lombardo} and with similar
conclusions. 
In short, 
the Dirac effect on the EoS of nucleonic matter is an essential saturating, and strongly  
density dependent, mechanism, which effectively accounts for the class of three-body forces
originating from virtual nucleon-antinucleon excitations. 
When hyperon degrees of freedom are included, for reasons of consistency, those should then be 
subjected to the same correction.            

We have incorporated DBHF effects in the present calculation, which amounts to involving the 
$\Lambda$ single-particle Dirac wave function in the self-consistent calculation through the 
$\Lambda$ effective mass, $m^*_{\Lambda}$.                             
Similarly to what is done for nucleons \cite{Mac89}, we fit 
the single-particle energy for $\Lambda$'s using the ansatz 
\begin{equation}
e_{\Lambda}(p) = \sqrt{(m^*_{\Lambda})^2 + p^2} + U_V^{\Lambda} 
\end{equation} 
with $m^*_{\Lambda}=m_{\Lambda}+U_S^{\Lambda}$, and $U_S^{\Lambda}$ and $U_V^{\Lambda}$ the scalar and vector potentials 
of the $\Lambda$ baryon. 

A problem with the        
J{\"u}lich NY potential in conjunction with DBHF calculations                             
is the use of the pseudoscalar coupling for the interactions of 
pseudoscalar mesons (pions and kaons) with nucleons and hyperons. For the reasons 
mentioned above (that is, the close relationship between Dirac effects and ``Z-diagram"
contributions), 
this relativistic correction is known to become unreasonably large when applied to 
a vertex involving pseudoscalar coupling. On the other hand, the gradient (pseudovector) 
coupling (also supported by chiral symmetry arguments) largely suppresses antiparticle contributions. 
To resolve this problem, one can make use of the on-shell equivalence between the pseudoscalar and
the pseudovector coupling, which amounts to 
relating the coupling constants as follows: 
\begin{equation}
g_{ps} = f_{pv}\frac{m_i + m_j}{m_{ps}},                   
\end{equation} 
where $g_{ps}$ denotes the pseudoscalar coupling constant and $f_{pv}$ the pseudovector one;
$m_{ps}$, $m_{i}$, and $m_{j}$ are the masses of the                         
pseudoscalar meson and the two baryons involved
in the vertex. 
This procedure can be made plausible by                                                               
writing down the appropriate 
one-boson-exchange amplitudes and observing that, redefining the coupling constants as above, we have 
(see Ref.~\cite{Mac89} for the two-nucleon case)                  
\begin{equation}
V_{pv} = V_{ps} + .....             
\end{equation} 
where the ellipsis stands for off-shell contributions. 
 Thus, the pseudoscalar coupling can be interpreted as pseudovector coupling where the 
off-shell terms are ignored. This is what we apply in our DBHF calculations.                  

In the coupled channel calculation, evaluation of $G_{N \Lambda}$ involves 
the transition potentials $V_{N \Lambda\rightarrow N \Lambda}$,
$V_{N \Lambda \leftrightarrow N \Sigma}$, and 
$V_{N \Sigma \rightarrow N \Sigma}$                                                   
(all with total channel isospin equal to 1/2), plus the corresponding exchange diagrams. 
Because in the present scenario (of a $\Lambda$ impurity in nucleonic matter)                                       
the $\Sigma$ hyperon is not given an effective mass,               
Dirac effects are applied         
only in $V_{N \Lambda \rightarrow N \Lambda}$. 
A diagram where not all of the baryon lines are Dirac-modified may yield a Dirac effect that is
artificially skewed. Moreover, since we find that the net contribution to the $\Lambda$ binding energy from the 
coupling to the $N \Sigma$ channels is rather small ($\approx$ 1.3 MeV), we anticipate
 Dirac effects from those channels to be negligibly small. 

Finally, a comment is in place concerning meson propagators. In standard DBHF calculations \cite{Mac89},
the so-called Thompson equation (a relativistic three-dimensional reduction of the 
Bethe-Salpeter equation) is used for two-baryon scattering. In the Thompson formalism, static
propagators           
are employed for meson exchange, i.e. 
\begin{equation}
-\frac{1}{m_{\alpha}^2 + ({\vec q}' - {\vec q})^2}  
\end{equation} 
where $m_{\alpha}$ denotes the mass of the exchanged meson and ${\vec q},{\vec q'}$ are 
the baryon momenta in their center-of-mass frame before and after scattering. 
The J{\"u}lich NY potentials are 
based upon time-ordered perturbation theory \cite{NY89} and use a meson propagator
given by 
\begin{equation}
\frac{1}{\omega_{\alpha} (z-E_i -E_j - \omega _{\alpha})} 
\end{equation} 
with $\omega _{\alpha}=\sqrt{m_{\alpha}^2 + ({\vec q'}-{\vec q})^2}$; 
$E_i=\sqrt{m_i^2 + {\vec q}'^2}$ and 
$E_j=\sqrt{m_j^2 + {\vec q}^2}$ are baryon energies and $z$ is the starting energy of the two-baryon 
system. In order to eliminate the energy dependence, Reuber {\it et al.} \cite{NY94} replaced the original
$z$ with 
\begin{equation}
z = \frac{1}{2} (m_1 + m_2 + m_3 + m_4),                                  
\end{equation} 
where the $m_i$'s denote the baryon masses of the four legs in the one-meson exchange diagram.
In any case, the J{\"u}lich meson propagator involves the baryon masses. Replacement of these
free-space masses with in-medium values would create medium effects on meson propagation which we do 
not wish to include in our nuclear matter calculations. The reason for keeping free-space masses in the 
meson propagator is twofold. First, standard DBHF calculations do not include medium effects
on meson propagation as they typically use Eq.~(7), which does not 
not depend on baryon masses. Second, medium effects on meson propagation constitute a separate class of 
effects that we are not concerned with in the present context  
and are typically not perceived as part of the DBHF approach.

Having taken the steps described above, 
we proceed to the DBHF calculation and 
find a moderate reduction of $B_{\Lambda}$, by approximately 4 MeV, due to Dirac effects.         
This is roughly 50\% of the corresponding effect on the nucleon potential. 
In Table II, contributions from selected partial waves are again shown.
A large part of the effect can be attributed to 
increased repulsion in $S$ and $P$ waves, especially 
$^3P_1$.

\begin{table}
\caption{Contributions to the $\Lambda$ binding energy from selected
partial waves obtained with the J{\"u}lich NY05 potential in a   
DBHF calculation.                  
} 
\begin{center}
\begin{tabular}{|c|c|}
%\hline
Partial wave & NY05, DBHF               \\                                         
                  \cline{1-2}
$^1S_0$ &           8.14                       \\ 
                  \cline{1-2}
$^3S_1$+$^3D_1$ & 36.45                      \\ 
                  \cline{1-2}
$^3P_0$   &    0.07                       \\ 
                  \cline{1-2}
$^1P_1$+$^3P_1$ &  -1.19                      \\ 
                  \cline{1-2}
$^3P_2$+$^3F_2$ &  3.21                       \\ 
\cline{1-2}
Total (all states)  &          47.4                       \\ 

%\hline
\end{tabular}
\end{center}
\end{table}

\section{Conclusions }                                
We have reported  on non-relativistic and                               
Dirac-Brueckner-Hartree-Fock predictions of the 
$\Lambda$ binding energy in nuclear matter at normal nuclear density.            
The magnitude of the Dirac effect
is approximately 1/2 of the corresponding effect on the binding of a nucleon in
nuclear matter. 

We also noticed and discussed the remarkably different predictions 
of this ``observable" as obtained using the 2005 or the 1994 versions
of the J{\"u}lich NY potentials. Given that there are noticeable differences
between the two free-space potentials, some potential model dependence is to
be expected. In this case, though, it would be appropriate to say that 
those free-space differences are considerably ``amplified" in the nuclear matter
calculation. Notice that this comparison was done between predictions as obtained from conventional
(BHF) calculations. But the conclusions would be unimpacted by
the Dirac effect, which is {\it much} smaller than the differences originating from the use
of the two potential models. 

The natural extension of this preliminary calculation                                                 
will be a fully self-consistent DBHF calculation of $U_{N}$, $U_{\Lambda}$, 
and $U_{\Sigma}$ for diverse $N$, $\Lambda$ and $\Sigma$ concentrations. 
Such work is in progress.

\section*{Acknowledgments}
Support from the U.S. Department of Energy under Grant No. DE-FG02-03ER41270 is 
acknowledged. I am grateful to Johann Haidenbauer for providing the nucleon-hyperon
potential code and for useful communications. 
%\section{References}

\end{document}